# A robust modeling framework for energy analysis of data centers

Nuoa Lei[†]
Department of Mechanical Engineering
Northwestern University
Evanston IL USA
nuoalei2021@u.northwestern.edu

## ABSTRACT

Global digitalization has given birth to the explosion of digital services in approximately every sector of contemporary life. Applications of artificial intelligence, blockchain technologies, and internet of things are promising to accelerate digitalization further. As a consequence, the number of data centers, which provide the services of data processing, storage, and communication services, is also increasing rapidly. Because data centers are energy-intensive with significant and growing electricity demand, an energy model of data centers with temporal, spatial, and predictive analysis capability is critical for guiding industry and governmental authorities for making technology investment decisions. However, current models fail to provide consistent and high dimensional energy analysis for data centers due to severe data gaps. This can be further attributed to the lack of the modeling capabilities for energy analysis of data center components including IT equipment and data center cooling and power provisioning infrastructure in current energy models. In this research, a technology-based modeling framework, in hybrid with a data-driven approach, is proposed to address the knowledge gaps in current data center energy models. The research aims to provide policy makers and data center energy analysts with comprehensive understanding of data center energy use and efficiency opportunities and a better understanding of macro-level data center energy demand and energy saving potentials, in addition to the technological barriers for adopting energy efficiency measures.

## 1 BACKGROUND AND RESEARCH MOTIVATION

Technologies such as Internet of Things, Blockchain, and Artificial Intelligence hold great promise for improving the efficiency of residential and industrial energy systems. Data centers form the backbone of such strategies, so understanding and managing their energy use is a critical component of sustainability moving forward. Currently, most macro-level data center energy modeling studies rely on simple extrapolations of outdated historical data to estimate the energy consumption of data centers [1–3,15], an approach which leads to unreliable estimation and creates barriers for effective policy making. Meanwhile, the booming of the big data era is leading to growing concern over the future scale of data center electricity use and its corresponding greenhouse gas emissions.

Despite the growing importance of data center energy use, only a few studies comprise the current credible quantitative literature on worldwide data center energy use. The first credible global study appeared in the year 2008 [9]. It focused on the period 2000 to 2005, which coincided with a rapid growth period in the history of the internet. Over these five years, the worldwide energy use of data centers was estimated to have doubled from 70.8 TWh/year to 152.5 TWh/year, with the latter value representing 1% of global electricity consumption. A subsequent bottom-up study [8], appearing in 2011, estimated that growth in global data center electricity use slowed from 2005 to 2010 due to steady technological and operational efficiency gains over the same period. According to this study, global data center energy use rose to between 203 TWh/year and 272 TWh/year by 2010, representing a 30%-80% increase compared to 2005. The latest global bottom-up estimates [12] produced a revised, lower 2010 estimate of 194 TWh/year, with only modest growth to around 205 TWh/year in 2018, or around 1% of global electricity consumption. The 2010 to 2018 flattening of global data center energy use has been attributed to substantial efficiency improvements in servers, storage devices, and network switches, and shifts away from traditional data centers toward cloud- and hyperscale-class data centers with higher levels of server virtualization and lower power usage effectiveness (PUE) values [11–13].

The scarcity of credible global data center energy analysis in the current literature can be attributed to the intensive input data requirements featured by current bottom-up models, which require a great amount of non-public data to generate reliable data center energy estimations. This could be further ascribed to the current deficiencies in modeling of the power use of data center IT equipment, and the power use of cooling and power provisioning system.

## 2 RESEARCH OBJECTIVE

Thus, the main goal of this research is to overcome current knowledge barriers by developing a new modeling framework that can quantify the direct energy use of worldwide data centers with temporal and spatial resolution from a bottom-up approach, where the bottom-up approach disaggregates the analysis of the complex data center energy system into energy analysis of its subsystems. This model will be able to analyze the power use of IT equipment based on different equipment features and technologies (e.g. CPU cores, server storage capacity, chip level cooling technologies, and thermal characteristics of interface materials). This model will also be able to analyze the power use of data center cooling and power provisioning system based on data center location, and data center cooling system options. Subsequently, this model will be able to predict time-series data center electricity use under uncertainty with spatial resolution (i.e. data enter type, and geographical locations). Furthermore, this modeling framework could decompose the changes in data center electricity consumption to the changes in the underlying contributors over time, providing insights to policy makers so that they can target specific energy use drivers for improving efficiency moving forward.

This modeling framework will provide a useful decision-making tool for policy makers, data center operators, and researchers in different fields. For policy makers, this modeling framework can provide reasonable data center energy projections and robust scenario analysis over temporal and spatial scales. This can assist decision-making in policy initiatives aimed at long-term macro-level data center energy reductions and effective power deployment for data centers. For data center operators, the modeling framework offers a credible methodology to estimate the benefits of adopting data center energy-saving strategies such as control system implementations and energy-efficient equipment replacement and helps them tailor investment strategies and apply these strategies with greater confidence. And for data center researchers, the modeling framework can provide a quantitative understanding of the key technological barriers hindering the energy saving potentials of data centers. This is helpful to precisely target areas which could lead to the greatest saving potentials in the future and to accelerate development in sustainable data centers.

## 3   RESEARCH SCOPE, METHODOLOGY, AND RATIONALE

The modeling framework will consist of 4 parts: (1) energy analysis of IT equipment including servers, storage, and network devices; (2) energy analysis of data center infrastructure (i.e. data center cooling and power provisioning systems); (3) bottom-up energy analysis with temporal and spatial resolution; and (4) data center energy index decomposition analysis. Each part of the modeling framework serves distinct roles in answering critical research questions related to the direct energy consumed by worldwide data centers, which are introduced as follows:

First, the energy analysis of IT equipment will be based on a hybrid modeling approach where part of the model structure is physics-based with the remaining being data-driven. Given various sources of uncertainty (i.e. parameter uncertainty, parametric variability, structural uncertainty) associated with the physics-based models, Bayesian model calibration will be used to tune the uncertain and unknown model parameters until the model predictions matches the observed data reasonably well [7] (data-driven part).

Second, the energy analysis of data center infrastructure was performed by developing thermodynamics-based PUE models, where the PUE is a value defined as the ratio of total data center power use to IT device power use [6]. The PUE models take location-specific climate conditions, data center energy system parameters (i.e. equipment specifications, system operational efficiency metrics, and indoor environment set points), and data center economizer choices as inputs for reliable PUE predictions. The PUE values in this research were predicted under uncertainty and verified by reported values from real data centers. In addition, a sub-module of this PUE model could calculate the achievable PUE values (optimal PUE based on state-of-the-art technologies) under different weather conditions through an integrated optimization algorithm, which is valuable for data center policy making concerning PUE regulations and baseline setting.

Third, the bottom up approach [4] is employed for data center macro-level energy analysis, with intent to quantify past, current, and near future electricity used by worldwide data centers, which takes a general form as described by Eq (1). The bottom-up approach will be based on power used by servers, external storage devices, network devices, and infrastructure equipment, which are basically the model outputs from step (1) and (2).

$$E^{dc} = \sum_j \left( \sum_i E_{ij}^{server} + \sum_i E_{ij}^{storage} + \sum_i E_{ij}^{network} \right) PUE_j \quad (1)$$

where $E^{dc}$ is data center electricity demand (kWh/year); $E_{ij}^{server}$ is electricity used by servers of class i in space type j (kWh/year); $E_{ij}^{storage}$ is electricity used by external storage devices of class i in space type j (kWh/year); $E_{ij}^{network}$ is electricity used by network devices of class i in space type j (kWh/year); $PUE_j$ is power usage effectiveness of data center in space type j (kWh/kWh).

Finally, the data center energy index decomposition analysis [16] will be performed based on the time-series model outputs from (1), (2), and (3) to investigate how changes in the worldwide data centers energy consumption could be attributed to the changes in underlying energy use drivers such as data center activity effects from server workloads/data storage/IP traffic/heat generation, regional effects, equipment density effects, structure effects of changes in data center types, structure effects of changes in equipment types, and equipment energy intensity effects (i.e. equipment energy efficiency).

Based on the aforementioned steps, the worldwide energy consumption of data centers can be modelled and analyzed under uncertainty with temporal and spatial resolution. The model can provide retrospective data center energy estimations and support uncertain energy projections under various scenarios (with the potential data center electricity savings associated with different energy efficiency improvements). In addition, it can quantitatively explain the underlying factors that might be targets of technology and policy interventions through the integrated sensitivity analysis module and the index decomposition analysis. These capacities will help stakeholders at different levels make robust decisions related to technology development and policy promotion.

## 4 RECENT RESULTS

### 4.1 Data Center PUE

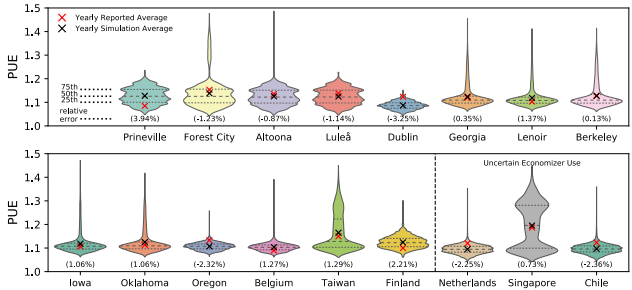

**Figure 1: Uncertainty quantification of location specific PUE values (Published in: [10])**

Figure 1 summarizes our recent analysis of data center PUE, based on the model described in step (2). Given the variabilities associated with weather conditions and energy system parameters for data centers, performing an uncertainty quantification enables one to assess confidence in the PUE model output. In our analysis, Monte Carlo simulation was used to predict the location-specific PUE values under uncertainties. Then the predicted values were compared with the reported PUE values to assess the validity of the proposed PUE models. To be more specific, 17 hyperscale data centers from Google and Facebook were chosen as the case studies for model validation. The results suggested that the model could generate acceptable PUE values for point estimations ($\pm 4\%$ relative errors at most) even though the uncertainty ranges of the yearly simulation results might be quite large. This level of accuracy benefits from applying location-specific climate data as model inputs, where the simulation average corresponds to the weighted average of the PUE simulations under the probabilistic weather condition in a specific location.

### 4.2 Worldwide Data Center Electricity Use

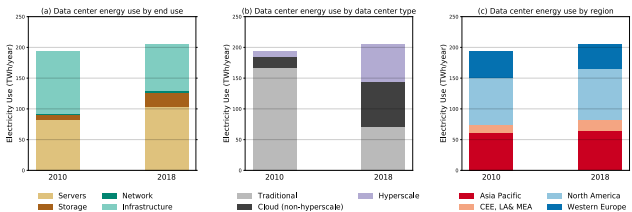

**Figure 2: Global data center energy consumption by end use, data center type, and region (published in: [12])**

Figure 2 summarizes our recent bottom-up data center energy estimates published in [12], which shows that global data centers were estimated to consume 194 TWh/year in 2010, with only modest growth to around 205 TWh/year in 2018. By 2018, the energy use of IT devices accounted for the largest share of data center energy use, due to substantial increases in the energy use of servers and storage devices driven by rising demand for data center computational and data storage services. The energy use of network switches is comparatively much smaller, accounting for only a small fraction over IT device energy use. In contrast, the energy use associated with data center infrastructure systems dropped significantly between 2010 and 2018, thanks to steady improvements in global average PUE values in parallel [5,14]. As a result of these counteracting effects, global data center energy use rose by only around 6% between 2010 and 2018, despite 11x, 6x, and 26x increases in data center IP traffic, data center compute instances, and installed storage capacity, respectively, over the same time period [12]. Figure 2 (b) summarizes data center energy use by major space type category, which shows that between 2010 and 2018, a massive shift away from smaller and less-efficient traditional data centers occurred, toward much larger and more efficient cloud data centers, and toward hyperscale data centers (a subset of cloud) in particular. Over this time period, the energy use of hyperscale data centers increased by about 4.5 times, while the energy use of cloud data centers (non-hyperscale) increased by about 2.7 times. However, the energy use of traditional data centers decreased by about 56%, leading to only modest overall growth in global data center energy use. As evident in Figure 1 (b), the structural shift away from traditional data centers has brought about significant energy efficiency benefits. Cloud and hyperscale data centers have much lower PUE values compared to traditional data centers, leading to substantially reduced infrastructure energy use. Moreover, cloud and hyperscale servers are often operated at much higher utilization levels (thanks to greater server virtualization and workload management strategies), which leads to far fewer required servers compared to traditional data centers. From a regional perspective, energy use is dominated by North America and Asia Pacific, which together accounted for around three-quarters of global data center energy use in 2018. The next largest energy consuming region is Western Europe, which represented around 20% of global energy use in 2018. It follows that data center energy management practices pursued in North America, Asia Pacific, and Western Europe will have the greatest influence on global data center energy use in the near-term.

## 5 CURRECT STATUS, NEXT STEPS, AND EXPECTED CONTRIBUTIONS

Currently, the student has accomplished the PUE modeling framework (step 2), and also contributed to the worldwide bottom-up data center energy recalibration published in [12] (step 3). The data center IT power model (step 1) is still under construction, but the modeling frameworks for Bayesian model calibration (step 1: second part) and index decomposition analysis (step 4) has been accomplished and will be submitted for publication in the next year.

The next steps for the student are to finish the physics-based IT power model, and then update the corresponding inputs for the data center bottom-up model and index decomposition analysis.

In general, the modeling framework is expected to: (1) provide a first-of-its-kind methodology to overcome knowledge barriers and deliver robust insights in light of data limitations and uncertainties about data center energy quantification with temporal and spatial resolution; (2) provide a flexible modeling framework that systematically integrates engineering analysis, physics, data science, and energy policy analysis so that it can be applied across different technologies, datasets, locations, and updated easily over time; (3) assist decision makers in understanding the important energy drivers of data centers and provide insight into critical technology barriers that hindering the energy saving potential of data centers; and (4) provide a useful tool to generate robust macro-level scenario comparisons over temporal and spatial scales in order to answer the questions of how technology improvement in the ICT industry that policy makers could help to reach long-term data center energy efficiency and energy saving goals.

## ACKNOWLEDGMENTS

The student would like to thank his PhD advisor, Professor Eric Masanet for his valuable guidance, and thank Leslie and Mac McQuown for the financial support of this research.

## REFERENCES

[1] Anders Andrae and Tomas Edler. 2015. On global electricity usage of communication technology: trends to 2030. *Challenges* 6, 1 (2015), 117–157.
[2] Lotfi Belkhir and Ahmed Elmeligi. 2018. Assessing ICT global emissions footprint: Trends to 2040 & recommendations. *J. Clean. Prod.* 177, (2018), 448–463.
[3] Peter Corcoran and Anders Andrae. 2013. Emerging trends in electricity consumption for consumer ICT. *Natl. Univ. Irel. Galway Connacht Irel. Tech Rep* (2013).
[4] Eric R. Masanet, Richard E. Brown, Arman Shehabi, Jonathan G. Koomey, and Bruce Nordman. 2011. Estimating the energy use and efficiency potential of US data centers. *Proc. IEEE* 99, 8 (2011), 1440–1453.
[5] International Energy Agency (IEA). 2017. *Digitalization and Energy*. IEA, Paris. Retrieved January 13, 2019 from https://www.iea.org/reports/digitalisation-and-energy
[6] Enrique Jaureguialzo. 2011. PUE: The Green Grid metric for evaluating the energy efficiency in DC (Data Center). In *2011 IEEE 33rd International Telecommunications Energy Conference (INTELEC)*, IEEE, 1–8.
[7] Marc C. Kennedy and Anthony O'Hagan. 2001. Bayesian calibration of computer models. *J. R. Stat. Soc. Ser. B Stat. Methodol.* 63, 3 (2001), 425–464.
[8] Jonathan Koomey. 2011. Growth in data center electricity use 2005 to 2010. *Rep. Anal. Press Complet. Req. N. Y. Times* 9, (2011), 161.
[9] Jonathan G. Koomey. 2008. Worldwide electricity used in data centers. *Environ. Res. Lett.* 3, 3 (2008), 034008.
[10] Nuoa Lei and Eric Masanet. 2020. Statistical analysis for predicting location-specific data center PUE and its improvement potential. *Energy* (2020).
[11] Eric Masanet and Nuoa Lei. 2020. HOW MUCH ENERGY DO DATA CENTERS REALLY USE? *Aspen Global Change Institute*. Retrieved from https://www.agci.org/sites/default/files/files%2Bqs/2020%20Q1%20Research%20Review_Masanet_Data%20Centers.pdf
[12] Eric R. Masanet, Arman Shehabi, Nuoa Lei, Sarah Smith, and Jonathan Koomey. 2020. Recalibrating global data center energy use estimates. *Science* 367, 6481 (February 2020).
[13] Arman Shehabi, Sarah Smith, Dale Sartor, Richard Brown, Magnus Herrlin, Jonathan Koomey, Eric Masanet, Nathaniel Horner, Inês Azevedo, and William Lintner. 2016. *United States Data Center Energy Usage Report*. Lawrence Berkeley National Lab (LBNL), Berkeley, CA (United States).
[14] Uptime Institute. 2018. *Uptime Institute Global Data Center Survey*. Retrieved October 25, 2018 from https://uptimeinstitute.com/2018-data-center-industry-survey-results
[15] Beth Whitehead, Deborah Andrews, Amip Shah, and Graeme Maidment. 2014. Assessing the environmental impact of data centres part 1: Background, energy use and metrics. *Build. Environ.* 82, (2014), 151–159.
[16] X. Y. Xu and B. W. Ang. 2014. Multilevel index decomposition analysis: Approaches and application. *Energy Econ.* 44, (2014), 375–382.